\documentclass{epl}

\usepackage{psfig}

\title{Coulomb and nuclear breakup of three-body halo nuclei}

\author{E. Garrido\inst{1} \and D.V. Fedorov\inst{2} \and A.S. Jensen\inst{2}}
\institute{
  \inst{1} Instituto de Estructura de la Materia, CSIC - Serrano 123, 
                                               E-28006 Madrid, Spain\\
  \inst{2} Institute of Physics and Astronomy, Aarhus University - 
                                              DK-8000 Aarhus C, Denmark 
}

\newcommand{\bd}[1]{ \mbox{\boldmath $#1$}  }

\begin{document}

\maketitle

\begin{abstract} 
We investigate dissociation reactions of loosely bound and spatially
extended three-body systems. We formulate a practical method for
simultaneous treatment of long-range Coulomb and short-range nuclear
interactions.  We use $^6$He (n+n+$\alpha$) and $^{11}$Li
(n+n+$^{9}$Li) as examples and study the two-neutron separation cross
sections as functions of target and beam energy.  Individual Coulomb
and nuclear as well as interference contributions are also extracted.
\end{abstract}

\section{Introduction} 

Nuclear halos were discovered by their surprisingly large reaction
cross sections with ordinary target 
nuclei~\cite{rii94,han95,tan96,jon98,ber93}.  Most of the information about
these unusual structures are obtained by detailed fragmentation
reaction studies~\cite{kob89,bla93,zin97,ale98,aum99}.  The understanding of
these halo systems is based on an effective clustering into few-body
structures reacting with the target. The main structure is essentially
agreed upon as two or three-body systems with $^{11}$Be (n+$^{10}$Be)
as the prototype of a two-body halo and $^6$He (n+n+$\alpha$) and
$^{11}$Li (n+n+$^{9}$Li) as three-body halos~\cite{zhu93}.

The reaction description is much less advanced and are mostly
available as fragmentary computations where one quantity is
investigated in one particular model with one set of 
parameters~\cite{ber93,oga92,suz94,ber98,ban98b,cob98}.  Systematic and
consistent calculations of many observables within one model are
highly desired but scarce~\cite{gar99}. The difficulties in an
accurate treatment are substantial, since even when the constituent
clusters and the target are oversimplifyingly considered to be
point-like, the reactions still involve three and four-body systems
for two and three-body halos, respectively. Approximations are
therefore inevitable.

One of the major problems in obtaining fragmentation cross sections is
to incorporate Coulomb and nuclear interactions in the same numerical
procedure~\cite{ber93,ber88}. This has been attempted for two-body
projectiles~\cite{das98,shy99} while three-body projectiles at best
are treated as effective two-body systems~\cite{ber93,ber88} or by
computing the Coulomb and nuclear contributions in independent 
models~\cite{ber93,ban98b,ber88,ber91,ban93}.  This situation is rather
unsatisfactory, since breakup reactions of nuclear halos are dominated
by nuclear and Coulomb interactions for light and heavy nuclear
targets, respectively~\cite{ber93,ber88,ban93}.

Significant improvements can be expected to be difficult, especially
for three-body projectiles, since a mixture of long-range and
short-range interactions is involved. The efforts may also be very
rewarding first by allowing the necessary treatment of specific
reactions and second due to the general nature of the problem and the
derived interest from other subfields of physics. We formulate in this
report a practical method to include Coulomb and nuclear interactions
simultaneously in investigations of breakup reactions of three-body
halo nuclei.

\section{Model assumptions}

We consider a three-body halo system colliding with a target. We
assume first that the intrinsic motion of the halo is slow compared to
the relative projectile-target motion. The three-body system does not
have time to adjust to the external field during the collision and the
sudden approximation therefore applies.  For point-like particles each
halo particle, called the participant, can then interact with the
target without disturbing the motion of the other two, called
spectators. We can then treat the reaction as independent collisions
and the total cross section is the sum of contributions from the three
participants.  Apart from energy and momentum conservation and
overlaps between initial and final states of the spectators we are
then left with three independent two-body problems.  The corresponding
interactions should then describe these two-body collisions to the
required level of accuracy. We use the phenomenological optical model
designed to describe elastic scattering and absorption from the
elastic channel.

The finite extension of both target and halo particles demands
additional considerations, since simultaneous collisions of more than
one halo particle then could be quite frequent in contradiction to our
basic assumption. Therefore in addition to the detailed treatment of
the essential contribution of the participant we use the simpler
``black sphere'' model to describe the smaller contributions from the
spectators. If the spectators are able to pass without touching the
target they are true spectators and otherwise they are counted as
absorbed by the target and consequently removed from the final
state. This is the optical model limit of very strong absorption for
short-range potentials. This division into detailed treatment of
participant and use of the black sphere for the spectators is only
meaningful when the halo is larger than the combined sizes of the
target and the participant. This is similar to the assumption used in
the formulation for a weakly bound projectile~\cite{ban67}.  A better
treatment of the collision between three-body halos and a target
almost inevitably has to deal with more than three-body configurations
in the final state or include properties of the intrinsic structure of
the halo particles and the target.

\section{Method}

The finite extension of the projectile particles and the target
destroy the clear division into participant and spectators, since the
spectators may hit the target in the same collision where the
participant is scattered or absorbed. We start by treating the
interaction between the target and the projectile constituents in the
black sphere model where the particle is absorbed inside a cylinder
with the axis along the beam direction and left untouched outside this
cylinder of a radius approximately equal to the target plus spectator
radius. If $P_c, P_{n1}$ and $P_{n2}$ are the probabilities for the
core, first and second neutron being inside this cylinder the reaction
probability is the sum of the probabilities of finding all three
constituents inside the cylinders ($P_{n1}P_{n2}P_c$), plus the
probability of finding two constituents inside (three terms like
$P_iP_j(1-P_k)$), plus the probability of finding only one constituent
inside the cylinder (three terms like $P_i(1-P_j)(1-P_k)$). We then
consider all possible combinations in the projectile-target
interaction. The sum of all these terms give the reaction probability,
that can be rewritten as $P_c + P_{n1}(1-P_c) +
P_{n2}(1-P_c)(1-P_{n1})$.  Each term in this probability vanishes
unless one particular projectile constituent is inside the cylinder.
This is the constituent chosen as participant, while the other two are
considered spectators. The optical model is then used for the
participant-target interaction.

\section{Participant treatment}

We consider collisions for an initial velocity $v$, the corresponding
momentum $p$, and total kinetic energy of $E A$, where $A$ is the mass
number of the projectile. The target (labeled 0) has charge $Z_0$ and
mass $m_0$, the participant (labeled $i$) has charge $Z_i$ and mass
$m_i$.  We label the spectators by $j$ and $k$, final state quantities
by primes, relative two-body coordinates and momenta between particles
$i$ and $k$ by $\bd{r}_{ik}$ and $\bd{p}_{ik}$ and between $i$ and the
center of mass of $j$ and $k$ by $\bd{r}_{i,jk}$ and
$\bd{p}_{i,jk}$. The initial wave function is a product of the
three-body wave function and a plane wave describing the relative
halo-target motion. The final state wave function is a product of
three terms, i.e. two distorted waves for the participant-target and
the spectator-spectator motion and a plane wave for the relative
motion of these two non-interacting two-body systems.  The distorted
waves are obtained by solving the Schr\"{o}dinger equation with the
appropriate two-body potentials.

The cross section of the participant $i$ has two contributions
corresponding to elastic scattering (diffraction) and absorption
(stripping).  The differential diffraction cross section reduces to a
factorized form when the participant has spin 0 or 1/2 and the target
has spin 0~\cite{gar99}
\begin{eqnarray}
&&  \frac{d^9\sigma _{el}^{(i)}(\bd{p}_{0i,jk}^{\prime },
\bd{p}_{jk}^{\prime },\bd{p}_{0i}^{\prime })}
{ d\bd{p}_{0i,jk}^{\prime } d\bd{p}_{jk}^{\prime } d\bd{p}_{0i}^{\prime } }
  =   \frac{d^3\sigma _{el}^{(0i)}(\bd{p}_{0i}
  \rightarrow  \bd{p}_{0i}^{\prime})} {d\bd{p}_{0i}^{\prime }} 
\nonumber \\ && \; \; 
\times  \left( 1 - |\langle \Psi| \exp(i \delta \bd{q}
 \cdot  \bd{r}_{i,jk}) |\Psi \rangle|^2 \right)
 |M_s(\bd{p}_{i,jk}, \bd{p}_{jk}^{\prime })|^2 \; ,  \label{eq5}
\end{eqnarray}
where $\bd{p}_{0i,jk}^{\prime } = \bd{p}_{i,jk} +
\bd{p}_{0}(m_j+m_k)/(m_0+m_i+m_j+m_k)$ in the rest frame of the
projectile, $\Psi$ is the initial three-body halo state, $\delta
\bd{q} = (\bd{p}_{i,jk}^{\prime } - \bd{p}_{i,jk})
(m_j+m_k)/(m_i+m_j+m_k)$ is the momentum transfer into the relative
participant-spectators motion and $M_s(\bd{p}_{i,jk},
\bd{p}_{jk}^{\prime })$ is the normalized overlap matrix element
between initial and final state spectator wave functions.

The first factor, $d^3\sigma _{el}^{(0i)}(\bd{p}_{0i} \rightarrow
\bd{p}_{0i}^{\prime}) / d\bd{p}_{0i}^{\prime }$, is the differential
cross section for the participant-target elastic scattering process.
It is obtained by numerical computation of the phase shifts including
simultaneously the nuclear and Coulomb potential.  At small momentum
transfers (large impact parameters) or large angular momenta the
Rutherford cross section is approached as the effect of the
short-range nuclear interaction then disappears.

The second factor in Eq.(\ref{eq5}) is constructed as one minus the
probability for staying in the ground state after transfer of the
momentum $\delta \bd{q}$ in the reaction, i.e. we remove the
probability for elastic scattering of the halo as a whole. This could
alternatively be done by orthogonalizing the final state in the
overlap matrix element to the initial three-body ground state.

Collisions at sufficiently large impact parameters only produce
virtual excitations corresponding to adiabatic 
motion~\cite{ber93,ber88}. The limiting impact parameter $b_a$ is determined
by equating the reaction time $2b_a/v$ with the time period $2 \pi
\hbar / B_{ps}$ in the relative motion of the participant-spectators
system with the corresponding binding energy $B_{ps}$. The Coulomb
interaction then transfers the momentum $q_a = Z_0Z_i e^2 p/(b_a A E)
= Z_0Z_i e^2 /(\pi c) \; B_{ps} /(\hbar c) \; (\gamma +1)\gamma^{-2}
\beta^{-2}$, where $\beta=v/c$ and $\gamma=1/\sqrt{1-\beta^2}$.

The energy transferred from target to participant, $ \delta E \equiv
\sqrt{{\bf p}_0^2 + m_0^2} - \sqrt{{\bf p_0^\prime}^2 + m_0^2}$, must
be larger than the three-body separation energy $B$. When ${\bf p_0}$
and ${\bf q} \equiv {\bf p_0} - {\bf p_0}^\prime$ are parallel $\delta
E$ is maximized. For this geometry we find for small $B$ compared to
the target rest mass that $\delta E = B$ implies that $q c \equiv q_L
c \approx B \sqrt{1+m_0^2c^2/p_0^2}$ which reduces to $B/v$ in the
non-relativistic limit. Thus $q$ must be larger than $q_L$ to produce
dissociation, but on the other hand dissociation is not the necessary
outcome for all $q > q_L$.  We exclude contributions from momentum
transfer smaller than the largest of $q_L$ and $q_a$.

The differential absorption (stripping) cross section, where the
participant in the sense of the optical model is absorbed by the
target, is obtained analogously~\cite{gar99}
\begin{eqnarray} \label{eq6}
 \frac{d^6\sigma _{abs}^{(i)}(\bd{p}_{0i,jk}^{\prime },\bd{p}_{jk}^{\prime })}
{ d\bd{p}_{0i,jk}^{\prime } d\bd{p}_{jk}^{\prime } }
=   \sigma _{abs}^{(0i)}(p_{0i}) \; 
 |M_s(\bd{p}_{i,jk}, \bd{p}_{jk}^{\prime })|^2 \; , 
\end{eqnarray}
where $\sigma _{abs}^{(0i)}$ is the participant-target absorption
cross section. The nine-dimensional differential cross section is now
reduced to six, since the absorbed or stripped particle inherently is
of no interest in the optical model description.  The factorizations
in both Eqs.(\ref{eq5}) and (\ref{eq6}) are incomplete, since
$\bd{p}_{0i}$ via momentum conservation is related to
$\bd{p}_{0i,jk}^{\prime }$ and $\bd{p}_{i,jk}$.

The total differential cross section for any reaction is now obtained
by adding the two contributions from scattering and absorption to
analogous contributions when particles $j$ and $k$ are the
participants. All these contributions are given on an absolute scale
and their relative weights are therefore determined in this model.  We
integrate over all unobserved momenta in Eqs.(\ref{eq5}) and
(\ref{eq6}).

\section{Spectator treatment.}

The spectator-target interaction is treated in the black sphere 
model. This model can be implemented by considering the distance $r_{ps}$
between the participant and the spectator~\cite{gar99}.
When $r_{ps}$ is small participant and spectator both would for short-range 
interactions interact with the target, and therefore according to
the black sphere model the spectator would be absorbed. The
distance $r_{ps}$ is then determining if the spectator is either
absorbed (distances smaller than $r_{ps}$) or undisturbed (distances 
larger than $r_{ps}$).
The absorption distances $r_{ps}$~\cite{gar99} are
related to the sizes of target and spectators and determined by
$\frac{3}{5} r_{ps}^2 = \langle r^2 \rangle_t + \langle r^2 \rangle_s
+ 2$ fm$^2$, where $\langle \rangle$ is the measured mean square
radius of target or spectator and 2 fm$^2$ is the square of the range
of the nucleon interaction. These radii $r_{ps}$ can also be obtained
from the parametrization in Eq.(28) of~\cite{gar99} with $r_0$ = 1.26
fm, 1.30 fm and 1.45 fm for the targets Pb, Cu and C, respectively.
\begin{figure}
\centerline{\psfig{figure=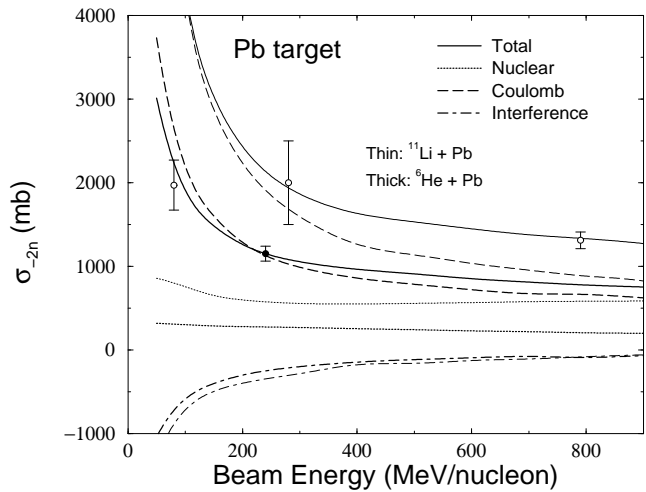,width=8.5cm,%
bbllx=3.4cm,bblly=1.9cm,bburx=19.7cm,bbury=23.4cm,angle=270}}
\vspace{0.2cm}
\caption{Two-neutron removal cross sections as functions of beam energy for
fragmentation of $^6$He (thick) and $^{11}$Li (thin) on a Pb-target.  We show
the total cross section (solid) as well as contributions from Coulomb
(dashed), nuclear (dotted) and interference terms (dot-dashed).  The curves
for $^6$He are lower than those of $^{11}$Li.  The experimental data are 
from~\protect\cite{kob89,bla93,zin97,aum99}, $^{11}$Li 
(open circles) and $^{6}$He
(filled circles).}
\label{fig1}
\end{figure}
In the same way when the charged participant is absorbed in the
optical model sense, it must have been close to the target and
close-lying spectators within the absorption distance must also be
counted as absorbed. On the other hand when the charged participant is
scattered small momentum transfer corresponds to a large impact
parameter.  These events only occur when the distance between
participant and target is large and it is then very unlikely that
spatially close-lying spectators are absorbed. All spectators are
therefore counted as scattered. For large momentum transfer in the
scattering process corresponding to impact parameters smaller than the
sum of target and participant radii the spectator is analogously
counted as absorbed.  For two-neutron dissociation cross sections this
division is irrelevant, since these contributions all are included in
the cross section.

\section{Results.}

We apply the procedure on the prominent nuclear three-body halos
$^6$He and $^{11}$Li. The neutron-neutron and the neutron-core
two-body interaction parameters are given in~\cite{cob98,gar99}.  The
optical model parameters are from~\cite{coo93} for the neutrons, 
from~\cite{nol87} for $\alpha$-particles and for $^{9}$Li-particles 
from~\cite{nol87} with range and diffuseness parameters 
from~\cite{zah96}. We furthermore drastically reduce the energy dependence
of the real potential in~\cite{nol87} to allow for the required huge
beam energy variation, i.e. $a_2 = -0.014$. The measured core-target
interaction cross sections are reproduced within error 
bars~\cite{bla93}.  The binding energy $B_{ps}$ between the $^{9}$Li and
$^4$He cores and the two neutrons must be introduced for the low
momentum cutoff. We use the scaling relation in~\cite{joh90} to obtain
$B_{ps}/B \approx 3$ for $^6$He and 1.4 for $^{11}$Li.
\begin{figure}
\centerline{\psfig{figure=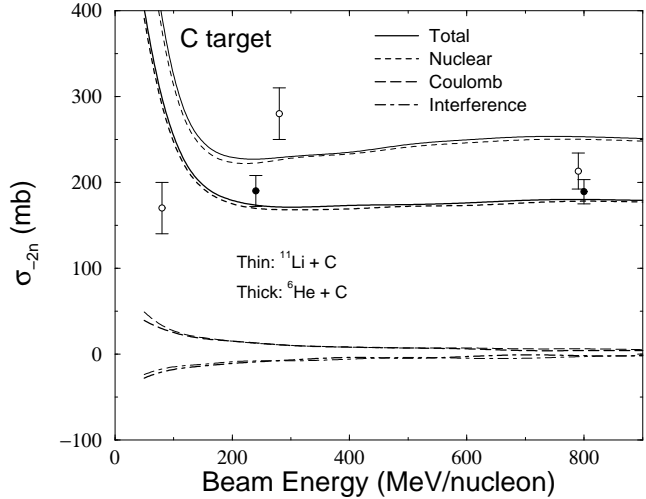,width=8.5cm,%
bbllx=3.4cm,bblly=2.4cm,bburx=19.7cm,bbury=23.4cm,angle=270}}
\vspace{0.2cm}
\caption[]{The same as Fig. \ref{fig1} for a Carbon target.  The experimental
data are from~\protect\cite{tan96,bla93,zin97,aum99}.}
\label{fig2}
\end{figure}
The two-neutron dissociation cross sections are shown in
fig.~\ref{fig1} as functions of beam energy for a Pb-target. The
theoretical uncertainties are mostly due to the inaccurate optical
model. Also the precise choice of cutoff parameter gives an
uncertainty on the Coulomb part, especially important at higher
energies. For Pb the adiabatic cutoff is the largest and therefore
decisive, i.e. $q_a > q_L$. The $^{11}$Li results are larger than
those of $^6$He because the Coulomb contribution roughly scales with
the square of the projectile charge and the nuclear part increases
with the projectile size. The small nuclear part has contributions
from both neutron and core as participants. They both decrease up to
about 200 MeV/nucleon then remaining roughly constant at higher
energies. Discussions of the individual behavior of the many
contributions are not possible here. The dominating Coulomb
contributions decrease with energy first strongly and then much slower
above 200 MeV/nucleon. The interference terms are negative and rather
small, but significant at low energies, where the total cross section
is below the Coulomb contribution.  The measured values are reproduced
at high energies but exceeded by a factor of two at low energy as
already noticed in~\cite{gar99}. Our results are consistent with
estimates reported in~\cite{oga92}.

The trends for the individual terms are roughly the same for C as for
Pb, see fig.~\ref{fig2}. The nuclear parts increase slowly with energy
above 200 MeV/nucleon.  The adiabatic cutoff is now the smallest and
therefore $q_L$ is decisive, i.e. $q_a < q_L$. We observe a small
increase of the total cross section with increasing energy.  The
nuclear contributions are completely dominating for a light target and
the numerical values are therefore also very similar for both
projectiles. Again we do not reproduce the (inconsistent) experimental
data at lower energies.

\begin{figure}
\centerline{\psfig{figure=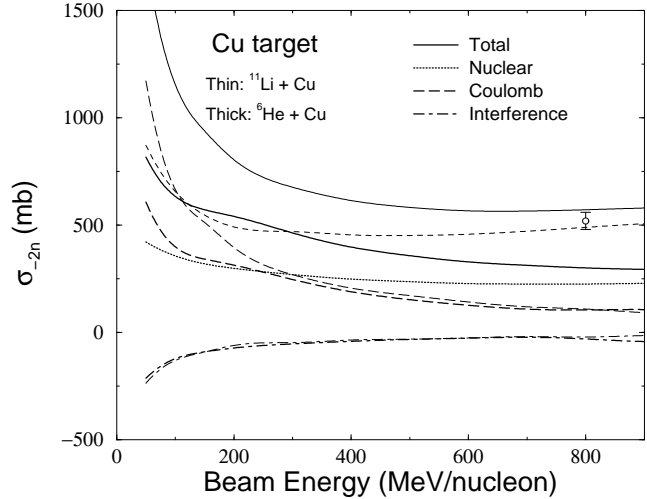,width=8.5cm,%
bbllx=3.4cm,bblly=2.4cm,bburx=19.7cm,bbury=23.4cm,angle=270}}
\vspace{0.2cm}
\caption[]{The same as fig.~\ref{fig1} for a Copper target. The experimental
point for $^{11}$Li is from~\protect\cite{kob89}.}
\label{fig3} 
\end{figure}

Coulomb and nuclear contributions are comparable for a medium heavy
target, see fig.~\ref{fig3}. The two cutoff parameters are now
comparable and $q_a$ is decisive at small energies whereas $q_L$
becomes larger at higher energies.  The trends for the individual
terms still remain the same as for C and Pb. Now the decreasing
Coulomb contribution crosses the almost energy independent nuclear
contribution at about 300 MeV/nucleon for $^6$He and at 200
MeV/nucleon for $^{11}$Li. The interference terms are still rather
small except at low energies.

\section{Conclusion.}

We have formulated a method to compute dissociation cross sections of
loosely bound three-body systems interacting with a mixture of short
and long-range potentials. The uncertainties in the numerical results
are dominated by the uncertainties in the two-body optical potentials
for light targets and in the cutoff parameter for heavy targets. The
already rather heavy computations seem to suffice as seen by comparing
with the measurements. The present consistent and systematic
calculations may prove useful as a guide for future experimental
investigations.  The dissociation cross sections are for example
predicted to increase as the beam energy decreases below 200
MeV/nucleon and slowly decrease or increase with increasing energy for
higher energies, respectively for heavy and light targets.

\acknowledgments
We thank K. Riisager for continuous 
discussions and suggestions.

\end{document}